\documentclass[11pt,a4paper]{article}
\pdfoutput=1
\usepackage{jheppub}
\usepackage{physics}

\newcommand{\liouv}{\mathcal{L}}
\newcommand{\cc}{\mathcal{C}}
\newcommand{\eave}[1]{\mathbb{E} \left[ #1 \right]}
\newcommand{\daga}{^\dag}

\newsavebox{\mstrut}
\newcommand{\bbra}[1]{
    \sbox{\mstrut}{\(#1\)}
    \mathinner{\left\langle\kern-0.5\ht\mstrut\left\langle{#1}\right|\mkern-2mu\right.}
}
\newcommand{\kett}[1]{
    \sbox{\mstrut}{\(#1\)}
    \mathinner{\left.\mkern-2mu\left|{#1}\right\rangle\kern-0.5\ht\mstrut\right\rangle}
}
\newcommand{\bbrakett}[2]{
    \sbox{\mstrut}{\(#1\)}
    \mathinner{\left\langle\kern-0.5\ht\mstrut\left\langle{#1}\right.\mkern-4mu\right.\left|\mkern-0mu\left.{#2}\right\rangle\kern-0.5\ht\mstrut\right\rangle}
}

\begin{document}

\title{Absence of operator growth for average equal-time observables in charge-conserved sectors of the Sachdev-Ye-Kitaev model}

\author[a,b]{Alessio Paviglianiti,}
\author[a,c]{Soumik Bandyopadhyay,}
\author[a,c]{Philipp Uhrich}
\author[a,c]{and Philipp Hauke}

\affiliation[a]{Pitaevskii BEC Center, INO-CNR and Dipartimento di Fisica, Università di Trento,
Via Sommarive 14, Trento I-38123, Italy}
\affiliation[b]{International School for Advanced Studies (SISSA), 
             via Bonomea 265, 34136 Trieste, Italy}
\affiliation[c]{INFN-TIFPA, Trento Institute for Fundamental Physics and Applications, Trento, Italy}

\emailAdd{apavigli@sissa.it}
\emailAdd{soumik.bandyopadhyay@unitn.it}
\emailAdd{philippjohann.uhrich@unitn.it}
\emailAdd{philipp.hauke@unitn.it}

\abstract{Quantum scrambling plays an important role in understanding thermalization in closed quantum systems. By this effect, quantum information spreads throughout the system and becomes hidden in the form of non-local correlations. Alternatively, it can be described in terms of the increase in complexity and spatial support of operators in the Heisenberg picture, a phenomenon known as operator growth. In this work, we study the disordered fully-connected Sachdev-Ye-Kitaev (SYK) model, and we demonstrate that scrambling is absent for disorder-averaged expectation values of observables. In detail, we adopt a formalism typical of open quantum systems to show that, on average and within charge-conserved sectors, operators evolve in a relatively simple way which is governed by their operator size. This feature only affects single-time correlation functions, and in particular it does not hold for out-of-time-order correlators, which are well-known to show scrambling behavior. Making use of these findings, we develop a cumulant expansion approach to approximate the evolution of equal-time observables. We employ this scheme to obtain analytic results that apply to arbitrary system size, and we benchmark its effectiveness by exact numerics. Our findings shed light on the structure of the dynamics of observables in the SYK model, and provide an approximate numerical description that overcomes the limitation to small systems of standard methods.}


\arxivnumber{2210.02427}

\maketitle
\flushbottom
	
\section{Introduction}\label{s:intro}
The investigation of out-of-equilibrium quantum systems is an extremely rich and active topic in modern statistical and condensed matter physics, which aims at clarifying fundamental concepts like thermalization and the emergence of the ensemble description~\cite{srednicki1994,polkovnikov2011,rigol2008,gogolin2016,dalessio2016,deutsch2018}. Theoretical advances in recent years have highlighted the importance of quantum information scrambling~\cite{sekino2008,iyoda2018,hosur2016} as a general feature associated to the relaxation of observables: Locally encoded information spreads throughout the system during the dynamics, and, while formally it is not lost, it cannot be recovered through any local measurement at late times. A paradigmatic example of a system exhibiting scrambling behavior is the Sachdev-Ye-Kitaev (SYK) model~\cite{kitaev2015,maldacena2016,gu2020,chowdhury2022}, which received much interest in recent research due to its relevance in multiple branches of physics, ranging from the study of black holes~\cite{kitaev2015,davison2017,kitaev2018,sachdev2019,sarosi2018} to non-Fermi liquids~\cite{sachdev2015,song2017,cha2020,sachdev2010}. There are multiple versions of the SYK model, parameterized by an even number $q$ and referred to as SYK$_q$. Each one consists of fermionic modes coupled via all-to-all disordered $q/2$-body interaction amplitudes. For $q\geq 4$, the model manifests quantum chaotic behavior~\cite{kitaev2015,polchinski2016,garciagarcia2018}, diagnosed by the presence of a quantum Lyapunov exponent in out-of-time-order correlators (OTOCs)~\cite{maldacena2016bis,hashimoto2017,kobrin2021}. Specifically, this exponent is found to saturate its theoretical upper bound~\cite{maldacena2016bis}, implying that quantum scrambling occurs as rapidly as possible. While multiple properties can be derived exactly in the thermodynamic limit~\cite{kitaev2015,sachdev2015,chowdhury2022}, studying the model at finite size remains a challenging yet essential problem, especially in view of future experimental investigations that have recently been proposed~\cite{danshita2017,garciaalvarez2017,pikulin2017,chew2017,chen2018,luo2019,wei2021}.

In this work, we investigate the disorder-averaged dynamics of observables in the SYK model, highlighting the manifestation of a counter-intuitive symmetry. In contrast to the scrambling nature of the system, we prove rigorously that disorder-averaged operators do not increase in complexity throughout the time-evolution. This corroborates the results of our previous work~\cite{bandyopadhyay2021}, which demonstrated numerically a non-trivial universality with respect to the choice of the initial state in the dynamics of some observables. The argument only applies to single-time observables, and, in particular, it does not hold for OTOCs: All correlators that can be used to diagnose scrambling and chaotic behavior are unaffected by our discussion, which resolves the apparent contradiction of our analysis with the well-known results from the literature. While our findings do not apply a priori to individual realizations of the model, they are expected to manifest in self-averaging single-time physical quantities for sufficiently large system sizes. In addition, working in a framework typical of open quantum systems, we leverage these results to show how to approximate the average dynamics by means of a cumulant expansion of the effective dynamical map. Specifically, the functions approximating the dynamics of observables can be computed analytically for arbitrary system sizes. Finally, we benchmark the performance of our approximation scheme through comparison with results found using exact diagonalization.

The paper is organized as follows. Section~\ref{s:model} presents the model and the superoperator framework adopted throughout the work. In Section~\ref{s:operator_size}, we introduce the concept of operator size, and we prove its conservation throughout the dynamics in ensemble average. Moreover, we discuss how this result does not conflict with well-known scrambling properties of the model. Then, in Section~\ref{s:cumulant_expansion} we present the cumulant expansion technique that we implement to approximate the dynamics. Our theoretical findings are tested in Section~\ref{s:application}, in which we investigate operator growth, and we probe the performance of the cumulant expansion using exact diagonalization simulations for the SYK$_4$ model. We summarize our findings in Section~\ref{s:conclusions}, and we discuss their possible applications, as well as future lines of research. Appendices expand the discussion of the main work, both by providing detailed descriptions of some proofs, and by showing explicit analytic results. 
	
\section{Model and focus of our study}\label{s:model}

We focus on the family of SYK models of $N$ complex fermion modes on a fully connected lattice, defined by the Hamiltonian
\begin{subequations}\label{hamiltonian}
\begin{equation}\label{hamiltonian-a}
    \hat{H} = \mathcal{K}_q \sum_{\substack{i_1<\dots<i_{q/2}\\j_1<\dots<j_{q/2}}} J_{i_1,\dots,i_{q/2};j_1,\dots,j_{q/2}} \hat{c}\daga_{i_1}\dots\hat{c}\daga_{i_{q/2}}\hat{c}_{j_1}\dots\hat{c}_{j_{q/2}},
    \end{equation}\begin{equation}
    \mathcal{K}_q = \sqrt{\frac{(q/2)!(q/2-1)!}{N^{q-1}}},
    \end{equation}
\end{subequations}
    where $q$ is an even integer number, and $J_{i_1,\dots,i_{q/2};j_1,\dots,j_{q/2}}$ are complex Gaussian independent random couplings with the following statistical properties ($\eave{...}$ denotes disorder averaging):
\begin{subequations}\label{disorder_properties}
\begin{align}
\label{disorder_properties-a}
\eave{J_{i_1,\dots,i_{q/2};j_1,\dots,j_{q/2}}}&=0,\\
\label{disorder_properties-b}
\eave{\left(J_{i_1,\dots,i_{q/2};j_1,\dots,j_{q/2}}\right)^2}&=0 \quad \text{if } \{i_1,\dots,i_{q/2}\}\neq\{j_1,\dots,j_{q/2}\},\\
\label{disorder_properties-c}
\eave{\left|J_{i_1,\dots,i_{q/2};j_1,\dots,j_{q/2}}\right|^2} &= J^2.
\end{align}
\end{subequations}
Notably, for $q>2$ the only integral of motion of any individual realization is the total charge $\hat{Q}=\sum_i \hat{n}_i$. 

Our analysis relies on the study of the dynamics of observable expectation values, namely
\begin{equation}
    \expval{\hat{W}(t)}  = \Tr\left(\hat{\rho}_0e^{i\hat{H}t} \hat{W} e^{-i\hat{H}t} \right),
\end{equation}
where typically (but not necessarily) $\hat{\rho}_0 = \ket{\psi_0}\bra{\psi_0}$ is a pure-state density matrix: This corresponds to preparing the system in a disorder-independent initial state $\ket{\psi_0}$, performing a quantum quench to the SYK$_q$ Hamiltonian, and tracking the time-evolution of physical quantities.
Specifically, since the Hamiltonian involves randomness, we focus on the disorder-averaged evolution $\eave{\expval{\hat{W}(t)}}$. A well-known feature of disordered systems is that disorder-averaging introduces mixing, making the average dynamics non-unitary~\cite{kropf2016}. As a consequence, it becomes natural to adopt a superoperator-based framework, typical of dissipative open quantum systems~\cite{minganti2018,manzano2020}. We introduce the Liouvillian $\liouv \bullet = -i\comm{\hat{H}}{\bullet}$, as well as the superoperators $\ell_\alpha \bullet = -i \comm{\hat{h}_\alpha}{\bullet}$, where $\alpha = \{i_1,\dots,i_{q/2};j_1,\dots,j_{q/2}\}$ is a multi-index, and $\hat{h}_\alpha = \hat{c}\daga_{i_1}\dots\hat{c}\daga_{i_{q/2}}\hat{c}_{j_1}\dots\hat{c}_{j_{q/2}}$ appears in the Hamiltonian of Eq.~\eqref{hamiltonian-a}. With this notation, the ensemble-averaged dynamics reads
\begin{equation}\label{observable_dyn}
    \eave{\expval{\hat{W}(t)}} = \Tr\left(\hat{\rho}_0 \eave{e^{-\liouv t}}\hat{W} \right).
\end{equation}
In what follows, we provide a detailed analysis of the time evolution of such disorder-averaged expectation values. 

\section{Operator size symmetry in the disorder-averaged ensemble}\label{s:operator_size}

In this section, we introduce the notion of operator size, and we show that it plays the role of a conserved quantum number for the average dynamics of operators. In this sense, when the total charge is fixed, no operator growth takes place for expectation values of observables. As we discuss, this result does not hold for more sophisticated correlators, e.g., OTOCs, which are able to diagnose quantum scrambling. 

\subsection{Operator size symmetry}\label{ss:size_sym}

An analysis of the spectral properties of the disorder-averaged time-evolution superoperator allows us to determine some exact features of the dynamics. Specifically, we will prove further below that the average dynamics distinguishes different operator sizes. Loosely speaking, we say that an operator consisting of the product of $m$ creation and $n$ annihilation operators has operator size $(m,n)$~\cite{carrega2021,roberts2018}. This definition is appropriate if all fermionic operators refer to different lattice sites, but we must analyze more carefully the cases involving one or more number operators $\hat{n}_i$. In any charge-conserved sector of the Hilbert space, the operator $\hat{Q}=\sum_i \hat{n}_i$ is proportional to the identity, and thus its size is $(0,0)$.~\footnote{It is worth noting that this size is assigned to the operator $\hat{Q}$ by definition. This follows naturally from the spectral structure of $\eave{e^{-\liouv t}}$, which regards the total charge as an operator with size $(0,0)$ (see Appendix~\ref{a:proof_op_size} for the details).} As a consequence, the size of the number operators $\hat{n}_i$ is not simply $(1,1)$, because each of them has a finite overlap with the identity operator. Despite this complication, it is possible to define the concept of operator size for the most general combination of creation and annihilation operators (see the discussion of Appendix~\ref{a:op_size}). Any linear combination of operators sharing a single common size also maintains that same size. This allows us to define an orthonormal basis of operators with well-defined sizes, over which any operator can be uniquely decomposed.

We now show that the dynamics preserves the sizes of operators on average, in the sense that it does not introduce components with operator sizes different from those present at $t=0$. We briefly present the main reasoning here, whilst a more detailed description of the following procedure can be found in Appendix~\ref{a:proof_op_size}. 

Taking the disorder average of a simple annihilation operator $\hat{c}_i(t)$ in the Heisenberg picture gives
\begin{equation}\label{expand_exp}
    \eave{e^{-\liouv t}}\hat{c}_i = \sum_{n=0}^\infty \frac{(\mathcal{K}_q t)^{2n}}{(2n)!}\eave{J_{\alpha_1}\dots J_{\alpha_{2n}}}\ell_{\alpha_1} \dots\ell_{\alpha_{2n}}\hat{c}_i,
\end{equation}
where the superoperators $\ell_{\alpha_i}$ are defined at the end of Section~\ref{s:model}, and the sum over repeated multi-indices is implicit. After using Wick's theorem, the previous equation results in a sum of nested commutators involving the operators $\hat{h}_{\alpha_1},\dots, \hat{h}_{\alpha_n}$, as well as their Hermitian conjugates $\hat{h}\daga_{\alpha_1},\dots, \hat{h}\daga_{\alpha_n}$, appearing in a certain order. In particular, each multi-index $\alpha_k$ is shared precisely by $\hat{h}_{\alpha_k}$ and $\hat{h}\daga_{\alpha_k}$, as follows from Eqs.~\eqref{disorder_properties-b} and~\eqref{disorder_properties-c}. If we expand each nested commutator in sequences of fermionic operators, for each $\hat{c}\daga_k$ appearing in any string there will always be exactly one $\hat{c}_k$, with the same index, somewhere within that same string.
For each chain of fermionic operators, we may use the anticommutation relations to move $\hat{c}_i$ towards one of the edges, for example the left one. Each new sub-string originated by this process involves only the operator $\hat{c}_i$ and multiple sums over all sites of paired creation and annihilation operators. Following this argument, any initial string is equal to the original operator $\hat{c}_i$ multiplied by a complex combination of fermionic operators that, however, cannot depend on any lattice site in particular, as all indices involved are summed over: As a consequence, the latter will be a function of the number of sites $N$ and the charge operator $\hat{Q}$. In conclusion, we have proved that
\begin{equation}\label{absence_growth}
    \eave{e^{-\liouv t}}\hat{c}_i = \hat{c}_i \hat{f}(t^2;N,\hat{Q}),
\end{equation}
where the dependence on a function of $t^2$ arises from the absence of odd terms in Eq.~\eqref{expand_exp}. The above result shows that, if we limit its action to a subset of the Hilbert space with fixed $Q$, $\hat{c}_i$ is an eigenoperator of the generator of the ensemble-averaged dynamics. In other words, $\hat{c}_i(t)$ manifests no operator growth (nor hopping) on average when the total charge is fixed. 

Our above argument is readily extended to all fermionic strings of type $\hat{S}_{m,n}=\hat{c}\daga_{i_1}\dots\hat{c}\daga_{i_m}\hat{c}_{j_1}\dots\hat{c}_{j_n}$ in which all indices are distinct, i.e.,
\begin{equation}\label{absence_growth_general}
    \eave{e^{-\liouv t}}\hat{S}_{m,n} = \hat{S}_{m,n} \hat{f}_{m,n}(t^2;N,\hat{Q}).
\end{equation}
These results apply also to strings containing repeated indices, i.e., number operators, and thus they hold for any generic operator with fixed size $(m,n)$: The proof is presented in Appendix~\ref{a:proof_op_size}. Notice that the functions $\hat{f}_{m,n}$ are independent of the specific lattice sites appearing in $\hat{S}_{m,n}$, consistently with the $S_N$ symmetry of the SYK model under ensemble average. 

In summary, we know how all operators with a well-defined size evolve on average: The operator size is conserved, and it determines the evolution in time. We point out that the derivation of this conclusion is model-independent, up to the requirement that the couplings in the Hamiltonian bear no dependence on lattice indices. For this reason, not only does this proof apply to any SYK$_q$ model, but also to linear combinations of SYK Hamiltonians with different values of $q$, as well as to other fully-connected homogeneous models. Our results apply also to the Majorana version of the SYK model~\cite{maldacena2016}, for which no charge operator $\hat{Q}$ can be defined; in that case, the functions $f_{m,n}$ will depend only on $t^2$ and $N$.~\footnote{This is seen by adapting the proof of Appendix~\ref{a:proof_op_size} to Majorana fermions. While for complex fermions the total charge emerges from the identity $\sum_i\hat{c}\daga_i\hat{c}_i = \hat{Q}$, for Majorana fermions $\hat{\psi}_i$ one has $\sum_i\hat{\psi}_i\hat{\psi}_i = N$.} In addition, we expect the result to extend also to fermionic and spin SYK-like models with non-Gaussian disorder, as well as bosonic versions of the system with Gaussian randomness.~\footnote{Gaussian disorder implies that the superoperators $\ell_\alpha$ are contracted in pairs. For fermionic systems, this is not fundamental, because the square of a creation or annihilation operator vanishes: Even if multiple operators with the same lattice index appear within a fermionic string, eventually the expression simplifies leaving at most one of them. A similar argument applies to spin models. In contrast, more-than-pairwise contractions generate new distinct operators for bosons, and the proof of operator size symmetry does not hold.}

\subsection{Implications for correlation functions and OTOCs}\label{ss:single_realizations}

The previous results might misleadingly be interpreted as total absence of operator growth in the SYK model, which would be an incorrect conclusion. We stress that operator size symmetry holds only under disorder average, whereas individual realizations of the system do clearly present operator growth (see Section~\ref{ss:operator_growth} for a detailed discussion). Indeed, operator growth is related to scrambling of quantum information, which is one key characteristic of the SYK model.

While it is important to distinguish physical features computed in single realizations and on average, there are cases in which the two coincide, namely for self-averaging observables. At suitably large $N$, such quantities approach their ensemble-averaged values for any single instance of the model, and thus operator size symmetry may manifest also for individual disorder realizations. Typical self-averaging quantities are expectation values of operators with extensive support~\cite{castellani2005,torresherrera2020}, which are precisely in the form of Eq.~\eqref{observable_dyn}. As a consequence, if, for instance, $\hat{W}=\hat{S}_{m,m}$ has a fixed operator size $(m,m)$, and $\hat{\rho}(0)$ has fixed number of particles $Q$, we have
\begin{equation}
    \expval{\hat{W}(t)} \approx \eave{\expval{\hat{W}(t)}} = f_{m,m}(t^2;N,Q)\Tr\left(\hat{W} \hat{\rho}_0\right),
\end{equation}
for large $N$. This implies that the functional form of the time-evolution is independent of the choice of the initial state, and the latter sets only the amplitude of the curve, as was observed in our previous work~\cite{bandyopadhyay2021}.

As anticipated previously, operator size conservation on average does not imply absence of scrambling in single physical realizations. An important aspect of the previous discussion is the limitation of our framework to the study of quantities that are linear in the time-evolution superoperator. In particular, OTOCs of the form
\begin{equation}
    \eave{\Tr\left(\hat{\rho}_\beta\comm{\hat{W}(t)}{\hat{V}(0)}^2\right)},
\end{equation}
which diagnose quantum chaotic behavior, are beyond the scope of our investigation, because they involve correlations between two superoperators $e^{-\liouv t}$, as well as the thermal density matrix $\hat{\rho}_\beta = e^{-\beta \hat{H}}/Z$. We conclude that quantum information scrambling must be encoded in these correlations, which arise because operators themselves are not self-averaging, namely $\hat{W}(t)\neq\eave{\hat{W}(t)}$ even if their quantum-mechanical expectation values become equal at large $N$. The same argument applies to the so-called Krylov complexity~\cite{parker2019,rabinovici2022,jian2021}, which is used to quantify operator growth. This is consistent with the belief that simple equal-time correlators are unable to capture the chaotic properties of the model, e.g., the quantum Lyapunov exponent. Such features are manifested only by more complicated quantities, such as, for instance, OTOCs.

To conclude this section, we point out that multiple studies in the literature focus on operator growth from the perspective of a distribution for the operator size~\cite{roberts2018,lensky2020,qi2019,lucas2020}. In such studies, operators are expanded over a basis of operators with fixed sizes, and a size distribution is obtained from the squared moduli of the expansion coefficients. Eventually, this framework allows one to compute the average size (see also~\cite{lin2019,jian2021}) and its dynamics, which can be directly related to OTOCs. Nevertheless, this bears no contradiction with our findings. The distribution has quadratic dependence on the time-dependent expansion coefficients, and thus its disorder-average probes correlations between two time-evolution superoperators; as discussed previously, our results do not apply to such a scenario.

\section{Cumulant expansion method}\label{s:cumulant_expansion}

In the previous section, we discussed the structure of the average time-evolution of operators. However, there is no straightforward way to explicitly determine the functions $\hat{f}_{m,n}(t^2;N,\hat{Q})$: The model is characterized by a single energy scale, and thus we cannot set up conventional perturbation theory starting from a known non-interacting solution. Instead, we now develop a cumulant expansion scheme to approximate the disorder-averaged dynamics of equal-time observables.

The numerical investigation presented in Ref.~\cite{bandyopadhyay2021} shows that the quench dynamics of an operator can be well approximated by a Gaussian, even though the agreement is not perfect. Motivated by this result, we look for an exponential representation of the disorder-averaged time-evolution superoperator, in such a way as to recover the Gaussian shape as the lowest order approximation, and to implement additional corrections. We formally write
\begin{equation}\label{cumulant_series}
    \eave{e^{-\liouv t}} = e^{\cc(t)} = \exp\left(\sum_{k=1}^\infty \frac{(Jt)^k}{k!}\cc_k\right),
\end{equation}
where $\cc(t)$ will be referred to as a cumulant generating superoperator, in agreement with usual nomenclature in statistics. Accordingly, the superoperators $\cc_k$ will be referred to as cumulants. Since we are considering Gaussian disorder, if $\liouv$ was just a scalar function instead of a superoperator, then all cumulants with $k>2$ would vanish. In the present case, however, the infinite series on the right-hand side of the previous equation does not terminate at finite order, because the operators that multiply different disordered couplings do not commute. 

A convenient way to determine the cumulants is by expanding the right-hand side of Eq.~\eqref{cumulant_series}, and comparing it to Eq.~\eqref{expand_exp} by matching equal powers of $t$. We immediately see that all superoperators $\cc_k$ with odd $k$ vanish. The identification of $\mathcal{C}_k$ requires the previous determination of all $\mathcal{C}_l$ with $l<k$, and thus the procedure is iterative. Here, we present the first two non-vanishing terms:
\begin{subequations}
\begin{align}\label{c2}
    \cc_2 &= \mathcal{K}_q^2 \sum_\alpha \ell_\alpha^2,\\
    \label{c4}
    \cc_4 &= \mathcal{K}_q^4 \sum_{\alpha,\beta}\left(\ell_\alpha\ell_\beta^2\ell_\alpha + \ell_\alpha\ell_\beta\ell_\alpha\ell_\beta-2\ell_\alpha^2\ell_\beta^2\right).
\end{align}
\end{subequations}
Leading order truncation of the cumulant expansion yields precisely the time-evolution superoperator used in Ref.~\cite{bandyopadhyay2021}, which was derived alternatively by writing an effective master equation for the ensemble-averaged density matrix. Higher-order cumulants provide corrections to the results of this previous work, allowing to achieve a better approximation of the dynamics.

For the SYK model, all cumulant superoperators commute. This property is proved by observing that each superoperator of type $\ell_{\alpha_1}\dots\ell_{\alpha_{2k}}$, in which the indices are contracted in pairs and summed over, preserves operators with well-defined size (see Appendix~\ref{a:proof_op_size} for the details), and thus all cumulants share the same eigenoperators. In particular, each individual cumulant fulfills an eigenvalue equation analogous to Eq.~\eqref{absence_growth_general}. As a consequence, we can differentiate Eq.~\eqref{cumulant_series} to obtain an effective master equation for the disorder-averaged density matrix in the Schrödinger picture,~\footnote{Notice that $e^{\cc(t)} = \eave{e^{-\liouv t}} = \eave{e^{\liouv t}}$ because all odd powers in the series expansion vanish.} yielding
\begin{equation}\label{me}
    \dv{}{t}\eave{\hat{\rho}(t)} = \dv{}{t}e^{\cc(t)}\hat{\rho}_0= \left(J\sum_{k=1}^\infty \frac{(Jt)^{2k-1}}{(2k-1)!}\cc_{2k}\right)\eave{\hat{\rho}(t)}.
\end{equation}
The sum within the round brackets is the effective Liouvillian that generates the disorder-averaged dynamics. 

For practical purposes, we can only evaluate the cumulant expansion up to some given finite order. Even if all $\cc_k$ were known, finding an analytic expression for $\cc(t)$ remains as hard as computing the exact disorder-averaged time-evolution superoperator; as a consequence, the series must be truncated. We point out, however, that even if only a finite number of terms is known, the resulting approximate description of the dynamics can be quite accurate over a wide time window. Indeed, using a truncated cumulant expansion scheme to characterize the average dynamics already proved successful in systems with non-Markovian noise~\cite{groszkowski2022}. Evaluating the series up to any given order is only as expensive as computing a short-time expansion up to that same order. Nevertheless, while the latter is limited to early times only, the former is potentially able to reasonably reproduce the evolution at arbitrary times. The reason lies in the exponential form of the time-evolution superoperator: Even when $\cc(t)$ is truncated, Eq.~\eqref{cumulant_series} still involves an infinite series of powers of $t$, and thus it can represent non-polynomial time-dependence. In addition, we can argue that the impact of higher-order cumulants is relevant only on longer timescales, as they are suppressed by a factor $(2k)!$.~\footnote{The timescale at which a cumulant becomes relevant is long if its magnitude in Eq.~\eqref{expand_exp} is small. Any $\cc_{2k}$ with $k>1$ should vanish if all superoperators $\ell_\alpha$ commuted, and thus each cumulant must contain a certain number $2M_{2k}$ of superoperators of type $\ell_{\alpha_1}\dots\ell_{\alpha_{2k}}$, half of them with the positive sign, half with the negative one (for example, see Eq.~\eqref{c4}): We do not expect $\cc_{2k}$ to grow as fast as the factor $(2k)!$, appearing at the denominator. On top of that, $M_{2k}$ is the number of irreducible Wick contractions arising from the $2k$th term of Eq.~\eqref{expand_exp}, which is smaller than $(2k-1)!!$ (it being the total number of contractions of order $2k$). By irreducible, we refer to those contractions that cannot be written as products of lower-order ones. For example, referring to Eq.~\eqref{c4}, the superoperators appearing with the plus sign are irreducible, whereas the other one is reducible. In conclusion, not only $2M_k<2(2k-1)!!<(2k)!$, but also different superoperators sum up in a destructive way.} Appendix~\ref{a:exact_evals} presents a numerical check of this statement, which indeed supports its validity. Suppose that the disorder-averaged density matrix quickly approaches a steady state, so that $\eave{\hat{\rho}(t)}\approx\eave{\hat{\rho}(\infty)}$ for $t \gtrsim\tau$. In this case, neglecting high-order cumulants that become relevant only at times greater than $\tau$ is a good approximation, because, as seen from Eq.~\eqref{me}, their action on the steady state density matrix is practically zero (notice that all cumulants share the same steady states, because they commute). In conclusion, for a system that thermalizes quickly, a finite number of cumulants is sufficient to obtain a valid approximation of the exact dynamics. As shown in Ref.~\cite{bandyopadhyay2021}, the quench dynamics of observables in the SYK model fall within this situation, as they manifest super-exponential relaxation to stationary values.

The above cumulant expansion allows us to approximate the functions $\hat{f}_{m,n}(t^2;N,\hat{Q})$. First, as observed previously, an equation analogous to Eq.~\eqref{absence_growth_general} can be written for each individual cumulant, so that
\begin{equation}
    \cc_{2k} \hat{S}_{m,n} = \hat{S}_{m,n}\hat{\lambda}^{(2k)}_{m,n}(N,\hat{Q}),
\end{equation}
where $\hat{S}_{m,n}$ is a generic operator in the form of Eq.~\eqref{size_operators}. The dynamical functions $\hat{f}_{m,n}(t^2;N,\hat{Q})$ are then expanded as
\begin{equation}\label{expand_fmn}
    \hat{f}_{m,n}(t^2;N,\hat{Q}) = \exp\left(\sum_{k=1}^\infty \frac{(J t)^{2k}}{(2k)!}\hat{\lambda}^{(2k)}_{m,n}(N,\hat{Q})\right).
\end{equation}
As argued before, a finite number of functions $\hat{\lambda}^{(2k)}_{m,n}$ can be sufficient to obtain a very good approximation of the exact result. The advantage of this approach is that eigenoperators of cumulants are known, and thus we can evaluate their eigenvalues analytically. Some exact expressions for the SYK$_4$ model are presented in Appendix~\ref{a:exact_evals}. From a practical point of view, this greatly simplifies the numerical application of our formalism: We can approximate Eq.~\eqref{expand_fmn} for arbitrarily large system sizes without resorting to exact diagonalization of cumulant superoperators.

\section{Numerical results}\label{s:application}

This Section presents numerical results that complement the previous theoretical findings. For a given operator, we first investigate the dynamics of those components  which have a well-defined size, and then we compare the exact time-evolution to the approximate one produced by the cumulant expansion. The exact dynamics is obtained by exact diagonalization of the SYK$_4$ Hamiltonian within a sector of fixed total charge $Q$. We evolve an initial disorder-independent pure state $\ket{\psi_0}$, and then we evaluate the expectation value $\expval{\hat{W}(t)}$ of an observable $\hat{W}$ that commutes with the charge $\hat{Q}$. In our numerics, we choose $\ket{\psi_0}$ to be the Néel state $\ket{1010\dots}$, where the ordering of lattice sites is arbitrary. Regarding the choice of the operator, we study the staggered magnetization $\hat{R} = \sum_{i=1}^N(-1)^{i+1}\hat{n}_i$ (the indices are assigned with the same convention used for the Néel state). Disorder-averaged quantities are computed by iterating this procedure multiple times for independent realizations of the disordered Hamiltonian given in Eq.~\eqref{hamiltonian-a}, and finally taking a statistical average over the sample.

\subsection{Operator size throughout the dynamics}\label{ss:operator_growth}

\begin{figure}[t]
\centering
    \includegraphics[width =0.7\linewidth]{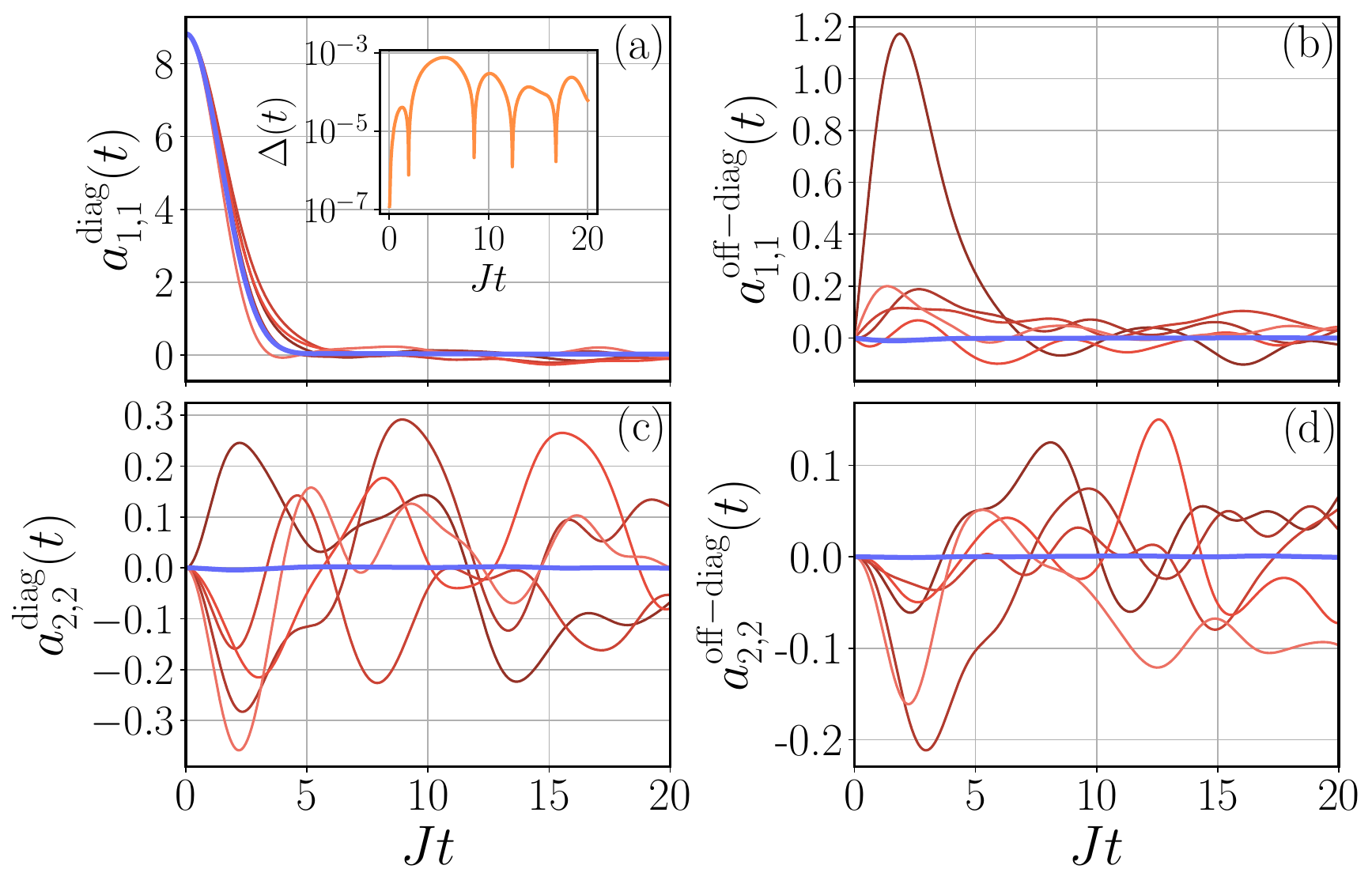}
    \caption{Dynamics of the coefficients $a^{(k)}_{m_k,m_k}(t)$ computed in single disorder realizations (red curves) and in ensemble average (blue curves) for the operator $\hat{R}$. The inset of panel (a) portrays the difference $\Delta(t) = \left|\eave{ a_{1,1}^\text{diag}(t)/a_{1,1}^\text{diag}(0)} - \eave{\expval{\hat{R}(t)}/\expval{\hat{R}(0)}}\right|$. Data is for $N=10$ at half filling $Q=N/2$. The ensemble averages are estimated by performing statistical averages over $10000$ independent disorder realizations. The terms of size $(2,2)$ vanish on average, in agreement with the predicted absence of operator growth. In contrast, there is a finite overlap with the diagonal operator of size $(1,1)$, and the time-evolution matches the full dynamics of $\expval{\hat{R}(t)}$, as can be appreciated by the inset of panel (a). Curves for individual realizations show that components with sizes different from $(1,1)$ are in general developed throughout the evolution.}
    \label{f:operator_growth}
\end{figure}

We now proceed to study the dynamics of individual components with well-defined size of a given operator. As mentioned previously, it is possible to define an orthonormal (with respect to the Hilbert-Schmidt product) basis of operators such that each basis element has a well-defined operator size. Using the notation introduced in Appendix~\ref{a:op_size}, we denote these basis operators as $\hat{T}^{(k)}_{m_k,n_k}$, and each of them has size $(m_k,n_k)$. We can extract their amplitudes in the expansion of $\hat{W}(t)$ by taking the Hilbert-Schmidt product
\begin{equation}\label{c_mn}
    a^{(k)}_{m_k,n_k}(t) = \Tr_Q\left(\left(\hat{T}^{(k)}_{m_k,n_k}\right)\daga \hat{W}(t)\right),
\end{equation}
where $\Tr_Q$ denotes the trace over states with fixed charge $Q$, as defined in Appendix~\ref{a:op_size}. For $\hat{W}=\hat{R}$, we use exact diagonalization to study the dynamics of some coefficients $a^{(k)}_{m_k,m_k}(t)$, representative of others. We focus on $m=1,2$, and for each operator size we design two basis operators, one diagonal and the other off-diagonal in the basis of Fock states. Following the prescription of Eq.~\eqref{size_operators} and fixing $Q=N/2$, we introduce
\begin{subequations}\label{test_operators}
\begin{align}
\label{test_operators-a}
\hat{T}_{1,1}^{(\text{diag})} &= \mathcal{N}^{(\text{diag})}_{1,1}\left(\hat{n}_1-\frac{1}{2}\right),
\\
\label{test_operators-b}
\hat{T}_{2,2}^{(\text{diag})} &= \mathcal{N}^{(\text{diag})}_{2,2}\left(\hat{n}_1\hat{n}_2-\frac{\hat{n}_1+\hat{n}_2}{2} + \frac{N}{4(N-1)}\right),
\\
\label{test_operators-c}
\hat{T}_{1,1}^{(\text{off-diag})} &= \mathcal{N}^{(\text{off-diag})}_{1,1}\left(\hat{c}\daga_1\hat{c}_2+\mathrm{h.c.}\right),
\\
\label{test_operators-d}
\hat{T}_{2,2}^{(\text{off-diag})} &= \mathcal{N}^{(\text{off-diag})}_{2,2}\left(\hat{c}\daga_1\hat{c}\daga_2\hat{c}_3\hat{c}_4+\mathrm{h.c.}\right),
\end{align}
\end{subequations}
where the prefactors $\mathcal{N}^{(k)}_{m_k,m_k}$ are such that these operators have unitary Hilbert-Schmidt norm. Figure~\ref{f:operator_growth} presents the disorder-averaged dynamics of the coefficients $a^{(k)}_{m_k,m_k}(t)$, as well as the evolution for some individual disorder realizations. As expected from the exact theoretical discussion in Section~\ref{ss:size_sym}, the operator $\hat{R}$ has no overlap with the size $(2,2)$. The coefficient $a^{(\text{off-diag})}_{1,1}(t)$ is also vanishing because the staggered magnetization only involves number operators $\hat{n}_i$. In contrast, $a^{(\text{diag})}_{1,1}(t)$ manifests a quick decay to zero, mirroring the super-exponential time-dependence of the expectation value $\expval{\hat{R}(t)}$ itself. This is substantiated by the inset in Figure~\ref{f:operator_growth}a, which shows that the time-dependence of $\eave{a_{1,1}^\text{diag}(t)}$ coincides with that of $\eave{\expval{\hat{R}(t)}}$ apart from fluctuations that can be attributed to the statistical error of the averages. If we consider the dynamics of a single realization of the SYK model, in general $\hat{R}(t)$ acquires components of size different from $(1,1)$, and it does not remain diagonal in the basis of Fock states. Still, Figure~\ref{f:operator_growth}a shows that the typical time-evolution is close to the disorder-averaged one. This effect is a consequence of self-averaging: In fact, we observe larger deviations from the average curve as we decrease the system size.

\subsection{Application of the cumulant expansion}\label{ss:cumulant_exp}

We now test the performance of the cumulant expansion by comparing it to the dynamics obtained through numerically exact simulations. As discussed in Section~\ref{s:operator_size}, each constituent of an operator $\hat{W}$ with well-defined size will evolve differently. The disorder-averaged dynamics thus reads
\begin{equation}\label{decompose_dynamics}
    \eave{\expval{\hat{W}(t)}} = \sum_m w_{m,m} f_{m,m}(t^2;N,Q),
\end{equation}
where 
\begin{equation}
    w_{m,m} = \Tr_Q\left(\hat{\rho}_0\mathcal{P}_{m,m}\hat{W}\right),
\end{equation}
and $\mathcal{P}_{m,m}$ is a projection superoperator over the subspace with operator size $(m,m)$. Notice that no components of size $(m,n)$ with $m\neq n$ can appear because $\hat{W}$ preserves the total charge. We explicitly see that the only role of the initial state is to set the amplitudes $w_{m,m}$. In our previous work~\cite{bandyopadhyay2021}, we provide numerical evidence that, for some observables, the time-dependence of $\eave{\expval{\hat{W}(t)}}$ manifests universality with respect to the choice of the initial state. This is readily explained by Eq.~\eqref{decompose_dynamics}: Operators that overlap with a unique operator size evolve according to a single dynamical function, and the initial state affects only the amplitude. In contrast, if two or more coefficients (excluding $w_{0,0}$) are non-zero, then changing $\hat{\rho}_0$ modifies the relative weights of different dynamical functions, and no universality is found. To summarize, the shape of the dynamics of an operator with well-defined size is independent of both the initial state and the precise definition of the operator itself.

\begin{figure}[t]
\centering
    \includegraphics[width=\linewidth]{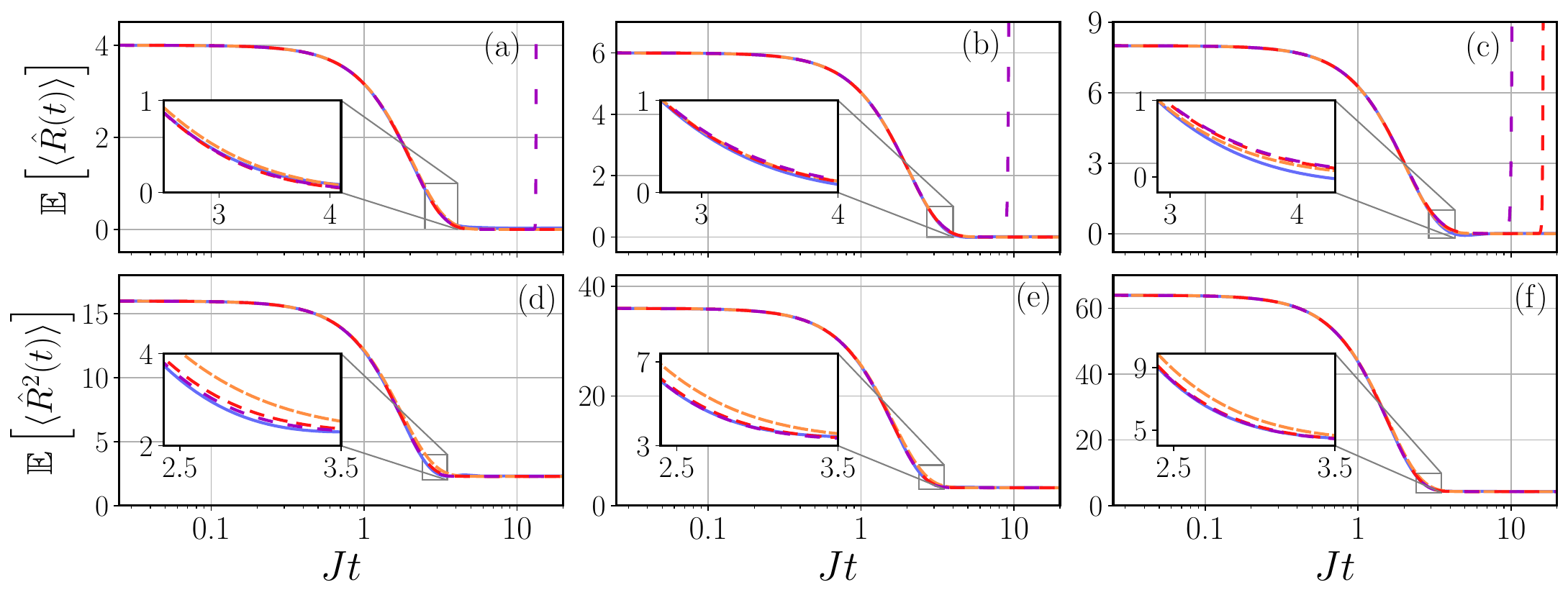}
    \caption{ Comparison of disorder-averaged dynamics of $\hat{R}$ and $\hat{R}^2$ obtained using exact diagonalization (solid blue curve), and the cumulant expansion truncated at second (dashed orange curve), fourth (dashed red curve), and sixth (dashed purple curve) order. Data is for $(\mathrm{a})$ and $(\mathrm{d})$ $N=8$ with 100000 samples, $(\mathrm{b})$ and $(\mathrm{e})$ $N=12$ with 3000 samples, $(\mathrm{c})$ and $(\mathrm{f})$ $N=16$ with 500 samples, at half filling $Q=N/2$.
    For $\hat{R}^2$, in all cases the addition of an extra cumulant manifestly improves the approximation of the exact dynamics (see insets). The same is true in the case of $\hat{R}$ when adding $\cc_4$ for $N=8,12$, whereas long-time divergences appear when adding $\cc_6$ for all considered system sizes, and already when adding $\cc_4$ for $N=16$. These pathologies are due to an inappropriate truncation of the cumulant at some system sizes. We also point out that generally (in the time windows preceding divergencies, if present) the variation induced by introducing the fourth cumulant is larger than that due to the sixth cumulant, which suggests quick convergence of the cumulant series.
    }
    \label{f:numerics}
\end{figure}

We study the dynamics of the staggered magnetization $\hat{R}$ and of its square $\hat{R}^2$. The former has operator size $(1,1)$, while the latter contains both sizes $(0,0)$ and $(2,2)$. Figure~\ref{f:numerics} presents the disorder-averaged dynamics of the operators for $N=8,12,16$, comparing it to the results of the cumulant expansion method truncated at different orders. Focusing on the operator $\hat{R}^2$, for all considered system sizes, the quality of the approximation is found to improve with the number of cumulants considered. In addition, the plots suggest that the curves converge rapidly to the exact result as more cumulants are included; even though this claim cannot be rigorously confirmed with only three cumulant superoperators, it agrees with the discussion of Section~\ref{s:cumulant_expansion}, where we argued that high-order corrections provide negligible contributions as compared to low-order ones. For the operator $\hat{R}$, instead, some curves manifest unphysical divergencies at late times. This happens because a cumulant eigenvalue $\lambda_{m,n}^{(2k)}$ can have a positive sign for some values of $N$ and $Q$. When this occurs, it is not possible to truncate the cumulant series to that specific order, as it would lead to divergent long-time behavior (see Eq.~\eqref{expand_fmn}). Instead, one should evaluate higher-order cumulants, until a negative eigenvalue is found, and the series can be truncated safely. We point out that also $\lambda_{2,2}^{(4)}$ and $\lambda_{2,2}^{(6)}$, which characterize the dynamics of $\hat{R}^2$, acquire positive values for larger system sizes than those considered here. Any divergence must be compensated by the presence of a higher-order cumulant with a negative eigenvalue that restores the relaxation to the steady state.

As previously exposed in Ref.~\cite{bandyopadhyay2021}, the steady state of the ensemble-averaged dynamics corresponds to the infinite temperature state. Some deviations are observed for small system sizes, but these can be attributed to finite-size effects. This result is reproduced by the cumulant expansion approach, as the identity is an eigenoperator of all cumulants, and is unique for all cases which we studied.

\section{Discussion and conclusions}\label{s:conclusions}
In this work, we have studied the out-of-equilibrium properties of the SYK model in terms of the spectral features of the effective time-evolution superoperator for the disorder-averaged dynamics. Owing to the absence of any spatial ordering, it is possible to identify the exact eigenoperators of $\eave{e^{-\liouv t}}$, and to prove that its (time-dependent) spectrum  manifests high degeneracies, corresponding to different operator sizes. As a result, for all system sizes, the dynamics is characterized by a non-trivial symmetry, which hides the scrambling properties of the model when probing the evolution of standard expectation values of observables in the ensemble average. This conclusion is limited to correlation functions that involve a single time-evolution superoperator, and thus it does not apply to multiple-time functions nor, specifically, OTOCs. Our findings prove rigorously that the dynamics of observables with well-defined operator size manifest universal features with respect to the initial state, as highlighted by our previous numerical study~\cite{bandyopadhyay2021}.

The proof of operator size symmetry implies the existence of dynamical functions, obtained from the diagonalization of $\eave{e^{-\liouv t}}$, that completely characterize the dynamics. Nevertheless, the demonstration does not indicate a simple way to compute them explicitly, and thus we developed a cumulant expansion scheme to achieve an approximation thereof. By construction, cumulant superoperators inherit the spectral structure of the exact time-evolution generator, and thus their eigenvalues can be determined analytically, as functions of the system size $N$ and total charge $Q$. A direct comparison between this approach and exact diagonalization simulations reveals that the method is successful in reproducing the evolution to good accuracy in all time regimes. In addition, numerical results corroborate theoretical arguments that the cumulant expansion converges quickly with the number of terms included. Although this approximation scheme appears to be very effective for the system considered here and, potentially, variants thereof, we expect its usefulness to be limited for generic models: In the absence of operator size symmetry, it is impossible to determine the eigenoperators of the cumulants analytically, which implies one would need to resort to computationally expensive explicit numerical diagonalization. This difficulty is overcome for the SYK model, thus enabling the cumulant expansion method to effectively compute the dynamical functions.

We believe our investigation enriches the current understanding of the SYK model by showcasing the presence of an unexpected symmetry, and by illustrating that scrambling can be completely absent in some physical quantities even if the model is quantum chaotic. While the main focus of our work is the complex SYK model, our findings directly extend to its Majorana counterpart, as argued in Section~\ref{ss:size_sym}. We expect our results to be relevant for future experimental implementations of the SYK model, as regular equal-time expectation values of observables are typical measurable quantities, further highlighting the need for measurement schemes that access more intricate observables~\cite{torlai2020}. In addition, the cumulant expansion approach that we developed manages to address arbitrary system sizes, thus enabling to access a regime that is out of reach of practical numerical simulations. We stress that a large enough system is fundamental to exploit the self-averaging property, which allows one to compare individual measurements to disorder-averaged quantities. 

A natural topic of future investigation is whether it is possible to generalize our open system framework to study OTOCs, and thus to potentially reveal generic features also for such quantities. Moreover, as mentioned previously, the present work bears implications for other systems: We believe that similar features to the ones discussed here are also manifested by other fully-connected disordered models. In particular, the proof of operator size symmetry does not rely on the details of the SYK model, and can be extended to fermionic and spin systems with other disorder distributions, as well as to bosonic models with Gaussian disorder.

\acknowledgments

We thank Andrea Legramandi for reading the manuscript and giving useful comments. This project has received funding from the European Research Council (ERC) under the European Union’s Horizon 2020 research and innovation programme (grant agreement No 804305), the Bundesministerium für Wirtschaft und Energie through the project ``EnerQuant” (project ID 03EI1025C), Provincia Autonoma di Trento, and by Q@TN, the joint lab between University of Trento, FBK-Fondazione Bruno Kessler, INFN-National Institute for Nuclear Physics and CNR-National Research Council. 
S.B. acknowledges CINECA for the use of HPC resources under ISCRA-C project ISSYK-2 (HP10CP8XXF).

\appendix

\section{Definition of operator size}\label{a:op_size}

As mentioned in Section~\ref{ss:size_sym}, strings of creation and annihilation operators acting on different lattice sites allow for an unambiguous definition of operator size. In contrast, number operators do not have a well-defined size, because they contain a contribution proportional to the total charge $\hat{Q}$, which is proportional to the identity when acting on any charge-conserved subspace. This argument suggests that the definition of operator size, in the most general case, should take into account the specific value of the charge $Q$ we are fixing. Therefore, throughout the following discussion, we limit the action of operators on kets with a fixed total charge $Q$.

The linear space of operators has the structure of a Hilbert space after introducing the Hilbert-Schmidt inner product
\begin{equation}
    \bbrakett{\hat{A}}{\hat{B}}=\Tr\left(\hat{A}\daga\hat{B}\right).
\end{equation}
Any operator $\hat{A}$ is represented as $\kett{\hat{A}}$ using a ket notation. Focusing on a specific charge sector, the overlap between two operators can be quantified through the charge-constrained Hilbert-Schmidt product, namely
\begin{equation}
    \bbrakett{\hat{A}}{\hat{B}}_Q = \Tr_Q\left(\hat{A}\daga\hat{B}\right),
\end{equation}
where $\Tr_Q$ denotes the trace over the Hilbert subspace with fixed charge $Q$. Notice that, in general, the charge-constrained trace does not have the cyclic property of the standard trace.

Having introduced the necessary tools, we now proceed to the definition of operators with fixed size. In order to understand the general case, it is instructive to first focus on those operators for which we are already able to provide a rigorous definition, namely strings of type $\hat{S}_{m,n}=\hat{c}\daga_{i_1}\dots\hat{c}\daga_{i_m}\hat{c}_{j_1}\dots\hat{c}_{j_n}$, where all indices are distinct. As discussed in Section~\ref{ss:size_sym}, such an operator has size $(m,n)$. Let $\hat{S}_{m',n'}$ be another operator of this type, with $(m,n)\neq(m',n')$. It is easily checked that these operators are orthogonal with respect to the charge-constrained Hilbert-Schmidt inner product
\begin{equation}\label{bbrakett}
    \bbrakett{\hat{S}_{m,n}}{\hat{S}_{m',n'}}_Q = 0.
\end{equation}
Equation~\eqref{bbrakett} can be used to generalize the notion of size to operators involving one or more $\hat{n}_i$. Specifically, we may extract a component with well-defined size from an operator of type $\hat{S}_{m,n}\hat{n}_{i_1}\dots\hat{n}_{i_k}$ by orthogonalizing it with respect to all operators of lower sizes. We now show how to perform this procedure in an iterative way.

Suppose that all operators with well-defined sizes $(m',n')$ with $m'\leq m$ and $n'\leq n$ are known. It is always possible to choose suitable linear combinations of them to obtain an orthonormal set. Let us denote the basis elements as $\hat{T}^{(k)}_{m_k,n_k}$, where $k$ is an index that counts them, and $(m_k,n_k)$ labels the size. We now pick a generic operator $\hat{S}_{m,n}$ with well-defined size $(m,n)$. Since the operator $\hat{n}_i$ is a combination of terms with sizes $(1,1)$ and $(0,0)$, the product $\hat{S}_{m,n}\hat{n}_i$ contains a component of size $(m+1,n+1)$, which is given by
\begin{equation}
\label{size_operators}
    \hat{S}_{m+1,n+1} = \hat{S}_{m,n}\hat{n}_i - \sum_k \bbrakett{\hat{T}^{(k)}_{m_k,n_k}}{\hat{S}_{m,n}\hat{n}_i}_Q \hat{T}^{(k)}_{m_k,n_k}.
\end{equation}
It is easily checked that, with this definition, $\hat{S}_{m+1,n+1}$ is orthogonal to all basis elements, and thus it does not belong to the manifold with operator sizes below or equal to $(m,n)$. We conclude that this operator has size $(m+1,n+1)$.

Proceeding in this way, we can formally determine all operators of size $(m+1,n+1)$ that are required, together with the trivial ones that do not involve any $\hat{n}_i$, to build an orthonormal set that spans all operators in the size sector $(m+1,n+1)$. Once these are known, we can iterate this procedure to generate operators with even larger sizes. Finally, replacing $Q\to\hat{Q}$ provides a general definition of $\hat{S}_{m,n}$, independent of the $Q$-sector. We point out that the actual computation of operators with large sizes can be quite expensive. Still, for practical purposes, one is typically interested in studying only operators with small sizes, where the implementation of the previous method is viable.

\section{Proof of operator size symmetry}\label{a:proof_op_size}
This Appendix complements the proof of operator size symmetry presented in Section~\ref{s:operator_size} by providing a more detailed derivation. Let us define
\begin{equation}
    \hat{A}_n = \eave{J_{\alpha_1}\dots J_{\alpha_{2n}}}\ell_{\alpha_1} \dots\ell_{\alpha_{2n}}\hat{c}_i\,,
\end{equation}
which is (up to a constant) of the form of the summands entering Eq.~\eqref{expand_exp}.  
We consider the first non-trivial term, namely
\begin{equation}\label{expanded_A1}
\begin{split}
    \hat{A}_1 &= \eave{J_{\alpha_1}J_{\alpha_2}}\ell_{\alpha_1} \ell_{\alpha_2}\hat{c}_i = -J^2\comm{\hat{h}_\alpha}{\comm{\hat{h}\daga_\alpha}{\hat{c}_i}}\\
    &= -J^2\left(\hat{h}_\alpha\hat{h}\daga_\alpha\hat{c}_i - \hat{h}_\alpha\hat{c}_i\hat{h}\daga_\alpha - \hat{h}\daga_\alpha\hat{c}_i\hat{h}_\alpha+\hat{c}_i\hat{h}\daga_\alpha\hat{h}_\alpha\right),
\end{split}
\end{equation}
where we made use of the disorder properties given in Eq.~\eqref{disorder_properties}. Each of the four terms in the previous equation can be expanded in strings of fermionic operators. For example, 
\begin{equation}
\begin{split}
    \hat{h}_\alpha\hat{c}_i\hat{h}\daga_\alpha &= \sum_{\substack{i_1<\dots<i_{q/2}\\j_1<\dots<j_{q/2}}}\left(\hat{c}\daga_{i_1}\dots\hat{c}_{j_{q/2}}\right)\hat{c}_i\left(\hat{c}\daga_{j_1}\dots\hat{c}_{i_{q/2}}\right)\\
    &=\frac{1}{(q/2)!^2}\sum_{\substack{i_1,\dots,i_{q/2}=1\\j_1,\dots,j_{q/2}=1}}^N\left(\hat{c}\daga_{i_1}\dots\hat{c}_{j_{q/2}}\right)\hat{c}_i\left(\hat{c}\daga_{j_1}\dots\hat{c}_{i_{q/2}}\right).
    \end{split}
\end{equation}
Similar expressions can be achieved for the other operators on the second line of Eq.~\eqref{expanded_A1}. For each of them, we can bring $\hat{c}_i$ to the left by using the anticommutation relations. In particular:
\begin{itemize}
    \item if there is an annihilation operator $\hat{c}_k$ on the immediate left of $\hat{c}_i$, we directly swap them, which yields a minus sign;
    \item if, instead, there is a creation operator $\hat{c}\daga_k$, we have $\hat{c}\daga_k\hat{c}_i = \delta_{i,k}-\hat{c}_i\hat{c}\daga_k$. The Kronecker delta restores the presence of $\hat{c}_i$ by constraining $\hat{c}_k$, present somewhere else on the fermionic string, to have $k=i$. Both terms originated by the anticommutation relation still involve a single $\hat{c}_i$ and sums of paired $\hat{c}\daga_k$ and $\hat{c}_k$ operators.
\end{itemize}
Eventually, the operator $\hat{c}_i$ is brought to the left of all fermionic strings. What remains on the right are homogeneous sums of operators, in the sense that they do not depend on the lattice index $i$, nor any other lattice index in particular: These can always be written in terms of the lattice size $N$ and the charge $\hat{Q}$. Finally, the discussion can be generalized to each operator $\hat{A}_n$, and thus Eq.~\eqref{absence_growth} follows.

The result we just proved for $\hat{c}_i$ is immediately generalized to any string of fermionic operators with distinct lattice indices, as defined above Eq.~\eqref{absence_growth_general}. In fact, we can repeat the same procedure for each  $\hat{c}^{(\dag)}_k$ independently, as all creation and annihilation operators anticommute; this leads to Eq.~\eqref{absence_growth_general}. In contrast, the validity of this result is not obvious for operators of the form given in Eq.~\eqref{size_operators}. We can, however, generalize the proof using induction. Consider a generic operator $\hat{S}_{m,n}$ of size $(m,n)$ (possibly involving also number operators), and assume that Eq.~\eqref{absence_growth_general} holds for all sizes below or equal to $(m,n)$. Under this inductive hypothesis, we want to prove that the eigenvalue equation is also valid for all operators of size $(m+1,n+1)$. Specifically, we need to show that the operator $\hat{S}_{m+1,n+1}$, defined as in Eq.~\eqref{size_operators}, also satisfies the eigenvalue equation. For this purpose, let $i$ and $j$ be lattice indices that do not appear in the definition of $\hat{S}_{m,n}$, and consider the action of the disorder-averaged time-evolution superoperator on $\hat{S}_{m,n}\hat{c}\daga_i\hat{c}_j$. We put no constraint on the values of $i$ and $j$ themselves, they may be equal or different. Throughout the procedure of moving the operators to the left, due to the appearance of Kronecker deltas when using the anticommutation relations, it is not guaranteed that $\hat{c}\daga_i$ and $\hat{c}_j$ remain in this order. In addition, some operators of sizes below or equal to $(m,n)$ may appear if $\hat{S}_{m,n}$ itself has been obtained through the orthogonalization procedure described in the previous section.~\footnote{Suppose that $\hat{S}_{m,n}$ is defined as in Eq.~\eqref{size_operators}, with $(m+1,n+1)\to(m,n)$. In this case, $\hat{S}_{m,n}\hat{c}\daga_i\hat{c}_j$ also contains terms of type $\hat{T}^{(k)}_{m_k,n_k}\hat{c}\daga_i\hat{c}_j$ with size $(m_k+1,n_k+1)$, where $m_k+1\leq m$ and $n_k+1\leq n$.} In general, we find
\begin{equation}\label{apply_superop_to_cc}
    \eave{e^{-\liouv t}}\left(\hat{S}_{m,n}\hat{c}\daga_i\hat{c}_j\right)  =\hat{S}_{m,n}\left(\hat{c}\daga_i\hat{c}_j\hat{A} + \hat{c}_j\hat{c}\daga_i\hat{B}\right)+\sum_k \hat{T}^{(k)}_{m_k,n_k}\left(\hat{c}\daga_i\hat{c}_j\hat{A}_k + \hat{c}_j\hat{c}\daga_i\hat{B}_k\right),
\end{equation}
where $\hat{A}$, $\hat{B}$, $\hat{A}_k$, and $\hat{B}_k$ are functions of $t^2$, $N$, and $\hat{Q}$, but not of $i$ nor $j$, and the operators $\hat{T}^{(k)}_{m_k,n_k}$ form an orthonormal basis for all sizes $(m',n')$ with $m'\leq m$ and $n'\leq n$, as introduced in Appendix~\ref{a:op_size}. For $i\neq j$, the operator $\hat{S}_{m,n}\hat{c}\daga_i\hat{c}_j$ is guaranteed to have size $(m+1,n+1)$. In particular, since $\hat{S}_{m,n}$ satisfies Eq.~\eqref{absence_growth_general} by assumption, then $\hat{S}_{m,n}\hat{c}\daga_i\hat{c}_j$ must fulfill the same equation (with $(m,n)\to(m+1,n+1)$) because we are simply adding unpaired creation and annihilation operators. Requiring that the eigenvalue equation is recovered for $i\neq j$ yields the conditions $\hat{A}-\hat{B}=\hat{f}_{m+1,n+1}$ and $\hat{A}_k=\hat{B}_k$. It follows that for $i=j$ we obtain
\begin{equation}\label{apply_superop_to_n}
    \eave{e^{-\liouv t}}\left(\hat{S}_{m,n}\hat{n}_i\right)  =\hat{S}_{m,n}\left(\hat{n}_i\hat{f}_{m+1,n+1} + \hat{B}\right) +\sum_k \hat{T}^{(k)}_{m_k,n_k}\hat{A}_k.
\end{equation}
Notice that $\hat{S}_{m,n}$ belongs to the space spanned by the basis elements $\hat{T}^{(k)}_{m_k,n_k}$, and thus we can absorb $\hat{S}_{m,n}\hat{B}$ in the last sum of Eq.~\eqref{apply_superop_to_n} by redefining the coefficients $\hat{A}_k$. For brevity, we do not change their notation, and we have
\begin{equation}\label{apply_superop_to_n_new}
    \eave{e^{-\liouv t}}\left(\hat{S}_{m,n}\hat{n}_i\right)  = \hat{S}_{m,n}\hat{n}_i\hat{f}_{m+1,n+1} +\sum_k \hat{T}^{(k)}_{m_k,n_k}\hat{A}_k.
\end{equation}

We will now relate $\hat{A}_k$ to the dynamical functions $\hat{f}_{m',n'}$. Adopting the bra-ket notation for operators, introduced in Appendix~\ref{a:op_size}, the action of the disorder-averaged time-evolution superoperator on the bra $\bbra{\hat{T}^{(k)}_{m_k,n_k}}$ is defined as
\begin{equation}\label{superop_bra}
    \bbra{\hat{T}^{(k)}_{m_k,n_k}} \eave{e^{-\liouv t}} = \left[\eave{e^{-\liouv t}}\kett{\hat{T}^{(k)}_{m_k,n_k}}\right]\daga = \bbra{\hat{T}^{(k)}_{m_k,n_k}\hat{f}_{m_k,n_k}}.
\end{equation}
We consider the following matrix element, which can be written in two ways by using Eqs.~\eqref{apply_superop_to_n_new} and~\eqref{superop_bra}:
\begin{equation}
    \begin{split}
    \bbra{\hat{T}^{(k)}_{m_k,n_k}}\eave{e^{-\liouv t}}\kett{\hat{S}_{m,n}\hat{n}_i}
   &= \bbrakett{\hat{T}^{(k)}_{m_k,n_k}}{\hat{S}_{m,n}\hat{n}_i\hat{f}_{m+1,n+1}} +\bbrakett{\hat{T}^{(k)}_{m_k,n_k}}{\hat{T}^{(k)}_{m_k,n_k}\hat{A}_k}\\
    &= \bbrakett{\hat{T}^{(k)}_{m_k,n_k}\hat{f}_{m_k,n_k}}{\hat{S}_{m,n}\hat{n}_i}.
\end{split}
\end{equation}
We have used the orthonormality condition of basis operators to obtain the first equality. Writing the Hilbert-Schmidt products explicitly, we obtain
\begin{equation}
    \Tr\left(\left(\hat{T}^{(k)}_{m_k,n_k}\right)\daga\hat{S}_{m,n}\hat{n}_i\left(\hat{f}_{m+1,n+1} -\hat{f}_{m_k,n_k}\right)\right)
    +\Tr\left(\left(\hat{T}^{(k)}_{m_k,n_k}\right)\daga \hat{T}^{(k)}_{m_k,n_k}\hat{A}_k\right) = 0,
\end{equation}
where we used $\hat{f}_{m,n}\daga = \hat{f}_{m,n}$,~\footnote{$\hat{f}_{m,n}$ involves only the identity and the charge operators, which are both Hermitian. In addition, it does not involve the imaginary unit, and thus it is Hermitian.} as well as the cyclicity of the trace.
Since $\hat{A}_k$, $\hat{f}_{m+1,n+1}$, and $\hat{f}_{m,n}$ all depend on $\hat{Q}$, the trace is conveniently decomposed as a sum of traces on subspaces with fixed charge $Q$, leading to
\begin{equation}\label{trace_eq}
    \sum_Q \bigg[\bbrakett{\hat{T}^{(k)}_{m_k,n_k}}{\hat{S}_{m,n}\hat{n}_i}_Q\left(f_{m+1,n+1}(Q)-f_{m_k,n_k}(Q)\right) +A_k(Q)\bigg] = 0.
\end{equation}
We now argue that each element of the sum over $Q$ must vanish individually. This is a consequence of charge conservation. Suppose we apply the disorder-averaged time-evolution superoperator to $\hat{S}_{m,n}\hat{n}_i\hat{g}$, where $\hat{g}=\hat{g}(\hat{Q})$ is an arbitrary function of the total charge operator. Charge conservation implies that $\hat{g}$ can be extracted from each commutator of Eq.~\eqref{expand_exp}, so that the analogue to Eq.~\eqref{apply_superop_to_n_new} reads
\begin{equation}
    \eave{e^{-\liouv t}}\left(\hat{S}_{m,n}\hat{n}_i\hat{g}\right)  =\hat{S}_{m,n}\hat{n}_i\hat{f}_{m+1,n+1}\hat{g}+\sum_k \hat{T}^{(k)}_{m_k,n_k}\hat{A}_k\hat{g}.
\end{equation}
Repeating the same calculations done previously results in a modified version of Eq.~\eqref{trace_eq}, in which additional weights $g(Q)$ appear in the sum. Since the function is arbitrary, each element of the sum must be zero, and we finally obtain
\begin{equation}
    A_k = \bbrakett{\hat{T}^{(k)}_{m_k,n_k}}{\hat{S}_{m,n}\hat{n}_i}_Q \left(f_{m_k,n_k}-f_{m+1,n+1}\right).
\end{equation}
Inserting this result in Eq.~\eqref{apply_superop_to_n} and reordering the terms, we finally arrive at
\begin{equation}
    \eave{e^{-\liouv t}}\hat{S}_{m+1,n+1} = \hat{S}_{m+1,n+1}\hat{f}_{m+1,n+1},
\end{equation}
where $\hat{S}_{m+1,n+1}$ is the operator of Eq.~\eqref{size_operators}. This proves the desired result.

\section{Analytic eigenvalues of cumulant superoperators}\label{a:exact_evals}
    Cumulant superoperators can be analytically diagonalized by applying them to fixed-size operators $\hat{S}_{m,n}$, and performing the procedure described in Appendix~\ref{a:proof_op_size} manually. The calculation can be carried out in an algorithmic way. Here, we present some exact results valid for the SYK$_4$ model, and for the operators of smallest sizes. For the first non-vanishing cumulant $\mathcal{C}_2$, we have:
\begin{subequations}
\begin{align}
    \begin{split}
    \hat{\lambda}^{(2)}_{1,0} &= -\frac{\hat{Q}(N-1)(N-\hat{Q}+1)}{N^3},
    \end{split}\\
    \begin{split}
    \hat{\lambda}^{(2)}_{1,1} &= -\frac{2(\hat{Q}-1)(N-\hat{Q}+1)}{N^2},
    \end{split}\\
    \begin{split}
    \hat{\lambda}^{(2)}_{2,1} &= -\frac{(N-1)\left[N(3\hat{Q}-2)-3\hat{Q}(\hat{Q}-1) \right]}{N^3},
    \end{split}\\
    \begin{split}
    \hat{\lambda}^{(2)}_{2,2} &= -\frac{2(N-1)\left[N(2\hat{Q}-3)-2\hat{Q}(\hat{Q}-2)\right]}{N^3}.
    \end{split}
\end{align}
\end{subequations}
We also present the exact result for one eigenvalue of the second non-vanishing cumulant
\begin{equation}
    \hat{\lambda}^{(4)}_{1,1} = \frac{2(\hat{Q}-1)\left[N^2(\hat{Q}-3) - 2N(\hat{Q}^2-\hat{Q}-3)\right]}{N^4} + \frac{2(\hat{Q}-1)\left[\hat{Q}(\hat{Q^2}+\hat{Q}-7)+1\right]}{N^4}\,.
\end{equation}
For higher-order cumulants, the calculation can be performed analogously but becomes increasingly cumbersome. All these expressions are rational functions of the number of lattice sites and the total charge, although this property is not used in any of the proofs in this article. Owing to Hermiticity of cumulants, we also have the property
\begin{equation}
    \hat{\lambda}^{(2k)}_{m,n}(N,\hat{Q}) = \hat{\lambda}^{(2k)}_{n,m}(N,\hat{Q}+m-n),
\end{equation}
and thus some eigenvalues are not independent. We numerically verified the exactness of some of the previous formulae (as well as some others that we do not provide here explicitly) by building matrix representations of the cumulant superoperators, and diagonalizing them in charge-conserved sectors of the Hilbert space. Specifically, we verified $\hat{\lambda}^{(2)}_{1,1}$, $\hat{\lambda}^{(2)}_{2,2}$, $\hat{\lambda}^{(4)}_{1,1}$, and $\hat{\lambda}^{(4)}_{2,2}$ for $N=4,6,8$, and also $\hat{\lambda}^{(6)}_{1,1}$ and $\hat{\lambda}^{(6)}_{2,2}$ for $N=4$.

\begin{figure}[t]
\centering
    \includegraphics[width=0.5\linewidth]{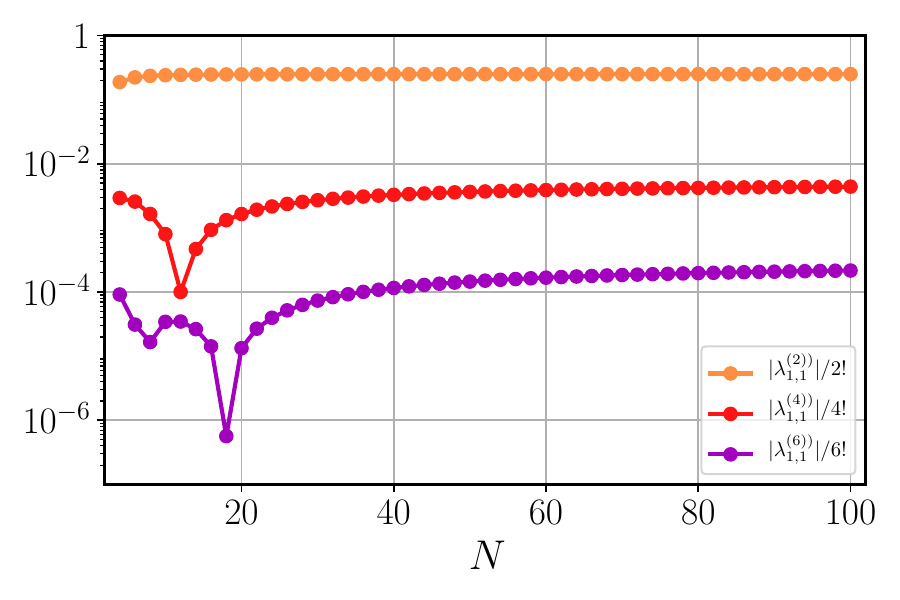}
    \caption{Comparison of $\left|\lambda_{1,1}^{(2k)}(N,N/2)\right|/{(2k)!}$ for the first three non-vanishing cumulant superoperators with $k=1,2,3$. For almost all values of $N$, we observe a clear separation of orders of magnitude. This suggests that truncating the cumulant expansion does not produce a dramatic error, as higher-order terms are suppressed more and more as $k$ increases.
    }
    \label{f:compare_cumulants}
\end{figure}

We may use the previous results to compare the magnitudes of different cumulants, and check if the contributions of high-order terms indeed become negligible, as argued in Section~\ref{s:cumulant_expansion}. Let us focus on the operator size of (1,1). Figure~\ref{f:compare_cumulants} compares $\left|\lambda_{1,1}^{(2k)}(N,N/2)\right|/{(2k)!}$ as functions of the system size $N$ and at half filling $Q=N/2$, for $k=1,2,3$. In agreement with previous arguments, different cumulants are characterized by distinct orders of magnitude, which decrease with the order $2k$. Note, however, that even though large powers of $t$ in Eq.~\eqref{expand_fmn} are suppressed by small prefactors, their presence can still be appreciable in the intermediate time regime describing the approach to the steady state, thus providing corrections to the Gaussian behavior found at lowest order. Despite the limited scope of the previous comparison, we believe similar results hold in general for other operator sizes, as well as higher-order cumulants. This supports the claimed effectiveness of the truncated cumulant expansion method in approximating the dynamics.

\bibliographystyle{JHEP}
\bibliography{syk}

\providecommand{\href}[2]{#2}\begingroup\raggedright\begin{thebibliography}{10}

\bibitem{srednicki1994}
M.~Srednicki, \emph{Chaos and quantum thermalization},
  \href{https://doi.org/10.1103/physreve.50.888}{\emph{Physical Review E}
  {\bfseries 50} (1994) 888–901}
  [\href{https://arxiv.org/abs/cond-mat/9403051}{{\ttfamily
  cond-mat/9403051}}].

\bibitem{polkovnikov2011}
A.~Polkovnikov, K.~Sengupta, A.~Silva and M.~Vengalattore, \emph{Colloquium:
  Nonequilibrium dynamics of closed interacting quantum systems},
  \href{https://doi.org/10.1103/RevModPhys.83.863}{\emph{Reviews of Modern
  Physics} {\bfseries 83} (2011) 863}
  [\href{https://arxiv.org/abs/1007.5331}{{\ttfamily 1007.5331}}].

\bibitem{rigol2008}
M.~Rigol, V.~Dunjko and M.~Olshanii, \emph{Thermalization and its mechanism for
  generic isolated quantum systems},
  \href{https://doi.org/10.1038/nature06838}{\emph{Nature} {\bfseries 452}
  (2008) 854–858} [\href{https://arxiv.org/abs/0708.1324}{{\ttfamily
  0708.1324}}].

\bibitem{gogolin2016}
C.~Gogolin and J.~Eisert, \emph{Equilibration, thermalisation, and the
  emergence of statistical mechanics in closed quantum systems},
  \href{https://doi.org/10.1088/0034-4885/79/5/056001}{\emph{Reports on
  Progress in Physics} {\bfseries 79} (2016) 056001}
  [\href{https://arxiv.org/abs/1503.07538}{{\ttfamily 1503.07538}}].

\bibitem{dalessio2016}
L.~D’Alessio, Y.~Kafri, A.~Polkovnikov and M.~Rigol, \emph{From quantum chaos
  and eigenstate thermalization to statistical mechanics and thermodynamics},
  \href{https://doi.org/10.1080/00018732.2016.1198134}{\emph{Advances in
  Physics} {\bfseries 65} (2016) 239–362}
  [\href{https://arxiv.org/abs/1509.06411}{{\ttfamily 1509.06411}}].

\bibitem{deutsch2018}
J.~M.~Deutsch, \emph{Eigenstate thermalization hypothesis},
  \href{https://doi.org/10.1088/1361-6633/aac9f1}{\emph{Reports on Progress in
  Physics} {\bfseries 81} (2018) 082001}
  [\href{https://arxiv.org/abs/1805.01616}{{\ttfamily 1805.01616}}].

\bibitem{sekino2008}
Y.~Sekino and L.~Susskind, \emph{Fast scramblers},
  \href{https://doi.org/10.1088/1126-6708/2008/10/065}{\emph{Journal of High
  Energy Physics} {\bfseries 2008} (2008) 065–065}
  [\href{https://arxiv.org/abs/0808.2096}{{\ttfamily 0808.2096}}].

\bibitem{iyoda2018}
E.~Iyoda and T.~Sagawa, \emph{Scrambling of quantum information in quantum
  many-body systems},
  \href{https://doi.org/10.1103/PhysRevA.97.042330}{\emph{Physical Review A}
  {\bfseries 97} (2018) 042330}
  [\href{https://arxiv.org/abs/1704.04850}{{\ttfamily 1704.04850}}].

\bibitem{hosur2016}
P.~Hosur, X.-L.~Qi, D.~A.~Roberts and B.~Yoshida, \emph{Chaos in quantum
  channels}, \href{https://doi.org/10.1007/jhep02(2016)004}{\emph{Journal of
  High Energy Physics} {\bfseries 2016} (2016) 4}
  [\href{https://arxiv.org/abs/1511.04021}{{\ttfamily 1511.04021}}].

\bibitem{kitaev2015}
A.~Kitaev, \emph{A simple model of quantum holography}, talks given at
  ``{Entanglement in Strongly-Correlated Quantum Matter},''
  \href{https://online.kitp.ucsb.edu/online/entangled15/kitaev/}{(Part 1,}
  \href{https://online.kitp.ucsb.edu/online/entangled15/kitaev2/}{ Part 2)},
  KITP (2015).

\bibitem{maldacena2016}
J.~Maldacena and D.~Stanford, \emph{Remarks on the {S}achdev-{Y}e-{K}itaev
  model}, \href{https://doi.org/10.1103/PhysRevD.94.106002}{\emph{Physical
  Review D} {\bfseries 94} (2016) 106002}
  [\href{https://arxiv.org/abs/1604.07818}{{\ttfamily 1604.07818}}].

\bibitem{gu2020}
Y.~Gu, A.~Kitaev, S.~Sachdev and G.~Tarnopolsky, \emph{Notes on the complex
  {S}achdev-{Y}e-{K}itaev model},
  \href{https://doi.org/10.1007/jhep02(2020)157}{\emph{Journal of High Energy
  Physics} {\bfseries 2020} (2020) 157}
  [\href{https://arxiv.org/abs/1910.14099}{{\ttfamily 1910.14099}}].

\bibitem{chowdhury2022}
D.~Chowdhury, A.~Georges, O.~Parcollet and S.~Sachdev,
  \emph{Sachdev-{Y}e-{K}itaev models and beyond: {W}indow into non-{F}ermi
  liquids}, \href{https://doi.org/10.1103/revmodphys.94.035004}{\emph{Reviews
  of Modern Physics} {\bfseries 94} (2022) 035004}
  [\href{https://arxiv.org/abs/2109.05037}{{\ttfamily 2109.05037}}].

\bibitem{davison2017}
R.~A.~Davison, W.~Fu, A.~Georges, Y.~Gu, K.~Jensen and S.~Sachdev,
  \emph{Thermoelectric transport in disordered metals without quasiparticles:
  The {S}achdev-{Y}e-{K}itaev models and holography},
  \href{https://doi.org/10.1103/physrevb.95.155131}{\emph{Physical Review B}
  {\bfseries 95} (2017) 155131}
  [\href{https://arxiv.org/abs/1612.00849}{{\ttfamily 1612.00849}}].

\bibitem{kitaev2018}
A.~Kitaev and S.~J.~Suh, \emph{The soft mode in the {S}achdev-{Y}e-{K}itaev
  model and its gravity dual},
  \href{https://doi.org/10.1007/jhep05(2018)183}{\emph{Journal of High Energy
  Physics} {\bfseries 2018} (2018) 183}
  [\href{https://arxiv.org/abs/1711.08467}{{\ttfamily 1711.08467}}].

\bibitem{sachdev2019}
S.~Sachdev, \emph{Universal low temperature theory of charged black holes with
  {AdS}$_{2}$ horizons}, \href{https://doi.org/10.1063/1.5092726}{\emph{Journal
  of Mathematical Physics} {\bfseries 60} (2019) 052303}
  [\href{https://arxiv.org/abs/1902.04078}{{\ttfamily 1902.04078}}].

\bibitem{sarosi2018}
G.~Sarosi, \emph{Ad{S}$_{2}$ holography and the {SYK} model},
  \href{https://doi.org/10.22323/1.323.0001}{\emph{Proceedings of XIII Modave
  Summer School in Mathematical Physics — PoS(Modave2017)} (2018) 1}
  [\href{https://arxiv.org/abs/1711.08482}{{\ttfamily 1711.08482}}].

\bibitem{sachdev2015}
S.~Sachdev, \emph{Bekenstein-{H}awking {E}ntropy and {S}trange {M}etals},
  \href{https://doi.org/10.1103/physrevx.5.041025}{\emph{Physical Review X}
  {\bfseries 5} (2015) 041025}
  [\href{https://arxiv.org/abs/1506.05111}{{\ttfamily 1506.05111}}].

\bibitem{song2017}
X.-Y.~Song, C.-M.~Jian and L.~Balents, \emph{Strongly {C}orrelated {M}etal
  {B}uilt from {S}achdev-{Y}e-{K}itaev {M}odels},
  \href{https://doi.org/10.1103/physrevlett.119.216601}{\emph{Physical Review
  Letters} {\bfseries 119} (2017) 216601}
  [\href{https://arxiv.org/abs/1705.00117}{{\ttfamily 1705.00117}}].

\bibitem{cha2020}
P.~Cha, N.~Wentzell, O.~Parcollet, A.~Georges and E.-A.~Kim, \emph{Linear
  resistivity and {S}achdev-{Y}e-{K}itaev ({SYK}) spin liquid behavior in a
  quantum critical metal with spin-1/2 fermions},
  \href{https://doi.org/10.1073/pnas.2003179117}{\emph{Proceedings of the
  National Academy of Sciences} {\bfseries 117} (2020) 18341–18346}
  [\href{https://arxiv.org/abs/2002.07181}{{\ttfamily 2002.07181}}].

\bibitem{sachdev2010}
S.~Sachdev, \emph{Strange metals and the {AdS}/{CFT} correspondence},
  \href{https://doi.org/10.1088/1742-5468/2010/11/p11022}{\emph{Journal of
  Statistical Mechanics: Theory and Experiment} {\bfseries 2010} (2010) P11022}
  [\href{https://arxiv.org/abs/1010.0682}{{\ttfamily 1010.0682}}].

\bibitem{polchinski2016}
J.~Polchinski and V.~Rosenhaus, \emph{The spectrum in the
  {S}achdev-{Y}e-{K}itaev model},
  \href{https://doi.org/10.1007/jhep04(2016)001}{\emph{Journal of High Energy
  Physics} {\bfseries 2016} (2016) 1–25}
  [\href{https://arxiv.org/abs/1601.06768}{{\ttfamily 1601.06768}}].

\bibitem{garciagarcia2018}
A.~M.~García-García, B.~Loureiro, A.~Romero-Bermúdez and M.~Tezuka,
  \emph{Chaotic-{I}ntegrable {T}ransition in the {S}achdev-{Y}e-{K}itaev
  {M}odel},
  \href{https://doi.org/10.1103/physrevlett.120.241603}{\emph{Physical Review
  Letters} {\bfseries 120} (2018) 241603}
  [\href{https://arxiv.org/abs/1707.02197}{{\ttfamily 1707.02197}}].

\bibitem{maldacena2016bis}
J.~Maldacena, S.~H.~Shenker and D.~Stanford, \emph{A bound on chaos},
  \href{https://doi.org/10.1007/jhep08(2016)106}{\emph{Journal of High Energy
  Physics} {\bfseries 2016} (2016) 106}
  [\href{https://arxiv.org/abs/1503.01409}{{\ttfamily 1503.01409}}].

\bibitem{hashimoto2017}
K.~Hashimoto, K.~Murata and R.~Yoshii, \emph{Out-of-time-order correlators in
  quantum mechanics},
  \href{https://doi.org/10.1007/jhep10(2017)138}{\emph{Journal of High Energy
  Physics} {\bfseries 2017} (2017) 138}
  [\href{https://arxiv.org/abs/1703.09435}{{\ttfamily 1703.09435}}].

\bibitem{kobrin2021}
B.~Kobrin, Z.~Yang, G.~D.~Kahanamoku-Meyer, C.~T.~Olund, J.~E.~Moore,
  D.~Stanford et~al., \emph{Many-{B}ody {C}haos in the {S}achdev-{Y}e-{K}itaev
  {M}odel},
  \href{https://doi.org/10.1103/physrevlett.126.030602}{\emph{Physical Review
  Letters} {\bfseries 126} (2021) 030602}
  [\href{https://arxiv.org/abs/2002.05725}{{\ttfamily 2002.05725}}].

\bibitem{danshita2017}
I.~Danshita, M.~Hanada and M.~Tezuka, \emph{{Creating and probing the
  Sachdev–Ye–Kitaev model with ultracold gases: Towards experimental
  studies of quantum gravity}},
  \href{https://doi.org/10.1093/ptep/ptx108}{\emph{Progress of Theoretical and
  Experimental Physics} {\bfseries 2017} (2017) }
  [\href{https://arxiv.org/abs/1606.02454}{{\ttfamily 1606.02454}}].

\bibitem{garciaalvarez2017}
L.~García-Álvarez, I.~L.~Egusquiza, L.~Lamata, A.~del Campo, J.~Sonner and
  E.~Solano, \emph{Digital {Q}uantum {S}imulation of {M}inimal {AdS/CFT}},
  \href{https://doi.org/10.1103/physrevlett.119.040501}{\emph{Physical Review
  Letters} {\bfseries 119} (2017) 040501}
  [\href{https://arxiv.org/abs/1607.08560}{{\ttfamily 1607.08560}}].

\bibitem{pikulin2017}
D.~I.~Pikulin and M.~Franz, \emph{Black {H}ole on a {C}hip: Proposal for a
  {P}hysical {R}ealization of the {S}achdev-{Y}e-{K}itaev model in a
  {S}olid-{S}tate {S}ystem},
  \href{https://doi.org/10.1103/physrevx.7.031006}{\emph{Physical Review X}
  {\bfseries 7} (2017) 031006}
  [\href{https://arxiv.org/abs/1702.04426}{{\ttfamily 1702.04426}}].

\bibitem{chew2017}
A.~Chew, A.~Essin and J.~Alicea, \emph{Approximating the
  {S}achdev-{Y}e-{K}itaev model with {M}ajorana wires},
  \href{https://doi.org/10.1103/physrevb.96.121119}{\emph{Physical Review B}
  {\bfseries 96} (2017) 121119}
  [\href{https://arxiv.org/abs/1703.06890}{{\ttfamily 1703.06890}}].

\bibitem{chen2018}
A.~Chen, R.~Ilan, F.~de~Juan, D.~I.~Pikulin and M.~Franz, \emph{Quantum
  {H}olography in a {G}raphene {F}lake with an {I}rregular {B}oundary},
  \href{https://doi.org/10.1103/physrevlett.121.036403}{\emph{Physical Review
  Letters} {\bfseries 121} (2018) 036403}
  [\href{https://arxiv.org/abs/1802.00802}{{\ttfamily 1802.00802}}].

\bibitem{luo2019}
Z.~Luo, Y.-Z.~You, J.~Li, C.-M.~Jian, D.~Lu, C.~Xu et~al., \emph{Quantum
  simulation of the non-{F}ermi-liquid state of {S}achdev-{Y}e-{K}itaev model},
  \href{https://doi.org/10.1038/s41534-019-0166-7}{\emph{npj Quantum
  Information} {\bfseries 5} (2019) 53}
  [\href{https://arxiv.org/abs/1712.06458}{{\ttfamily 1712.06458}}].

\bibitem{wei2021}
C.~Wei and T.~A.~Sedrakyan, \emph{Optical lattice platform for the
  {S}achdev-{Y}e-{K}itaev model},
  \href{https://doi.org/10.1103/physreva.103.013323}{\emph{Physical Review A}
  {\bfseries 103} (2021) 013323}
  [\href{https://arxiv.org/abs/2005.07640}{{\ttfamily 2005.07640}}].

\bibitem{bandyopadhyay2021}
S.~Bandyopadhyay, P.~Uhrich, A.~Paviglianiti and P.~Hauke, \emph{Universal
  equilibration dynamics of the {S}achdev-{Y}e-{K}itaev model}, {\emph{ArXiv}
  (2021) } [\href{https://arxiv.org/abs/2108.01718}{{\ttfamily 2108.01718}}].

\bibitem{kropf2016}
C.~M.~Kropf, C.~Gneiting and A.~Buchleitner, \emph{Effective {D}ynamics of
  {D}isordered {Q}uantum {S}ystems},
  \href{https://doi.org/10.1103/PhysRevX.6.031023}{\emph{Physical Review X}
  {\bfseries 6} (2016) 031023}
  [\href{https://arxiv.org/abs/1511.08764}{{\ttfamily 1511.08764}}].

\bibitem{minganti2018}
F.~Minganti, A.~Biella, N.~Bartolo and C.~Ciuti, \emph{Spectral theory of
  {L}iouvillians for dissipative phase transitions},
  \href{https://doi.org/10.1103/physreva.98.042118}{\emph{Physical Review A}
  {\bfseries 98} (2018) 042118}
  [\href{https://arxiv.org/abs/1804.11293}{{\ttfamily 1804.11293}}].

\bibitem{manzano2020}
D.~Manzano, \emph{A short introduction to the {L}indblad master equation},
  \href{https://doi.org/10.1063/1.5115323}{\emph{AIP Advances} {\bfseries 10}
  (2020) 025106} [\href{https://arxiv.org/abs/1906.04478}{{\ttfamily
  1906.04478}}].

\bibitem{carrega2021}
M.~Carrega, J.~Kim and D.~Rosa, \emph{Unveiling {O}perator {G}rowth {U}sing
  {S}pin {C}orrelation {F}unctions},
  \href{https://doi.org/10.3390/e23050587}{\emph{Entropy} {\bfseries 23} (2021)
  587} [\href{https://arxiv.org/abs/2007.03551}{{\ttfamily 2007.03551}}].

\bibitem{roberts2018}
D.~A.~Roberts, D.~Stanford and A.~Streicher, \emph{Operator growth in the {SYK}
  model}, \href{https://doi.org/10.1007/jhep06(2018)122}{\emph{Journal of High
  Energy Physics} {\bfseries 2018} (2018) 122}
  [\href{https://arxiv.org/abs/1802.02633}{{\ttfamily 1802.02633}}].

\bibitem{castellani2005}
T.~Castellani and A.~Cavagna, \emph{Spin-glass theory for pedestrians},
  \href{https://doi.org/10.1088/1742-5468/2005/05/p05012}{\emph{Journal of
  Statistical Mechanics: Theory and Experiment} {\bfseries 2005} (2005) P05012}
  [\href{https://arxiv.org/abs/cond-mat/0505032}{{\ttfamily
  cond-mat/0505032}}].

\bibitem{torresherrera2020}
E.~J.~Torres-Herrera, I.~Vallejo-Fabila, A.~J.~Mart\'{\i}nez-Mendoza and
  L.~F.~Santos, \emph{Self-averaging in many-body quantum systems out of
  equilibrium: Time dependence of distributions},
  \href{https://doi.org/10.1103/PhysRevE.102.062126}{\emph{Physical Review E}
  {\bfseries 102} (2020) 062126}
  [\href{https://arxiv.org/abs/2005.14188}{{\ttfamily 2005.14188}}].

\bibitem{parker2019}
D.~E.~Parker, X.~Cao, A.~Avdoshkin, T.~Scaffidi and E.~Altman, \emph{A
  {U}niversal {O}perator {G}rowth {H}ypothesis},
  \href{https://doi.org/10.1103/physrevx.9.041017}{\emph{Physical Review X}
  {\bfseries 9} (2019) 041017}
  [\href{https://arxiv.org/abs/1812.08657}{{\ttfamily 1812.08657}}].

\bibitem{rabinovici2022}
E.~Rabinovici, A.~S{\'{a}}nchez-Garrido, R.~Shir and J.~Sonner, \emph{Krylov
  complexity from integrability to chaos},
  \href{https://doi.org/10.1007/jhep07(2022)151}{\emph{Journal of High Energy
  Physics} {\bfseries 2022} (2022) 151}
  [\href{https://arxiv.org/abs/2207.07701}{{\ttfamily 2207.07701}}].

\bibitem{jian2021}
S.-K.~Jian, B.~Swingle and Z.-Y.~Xian, \emph{Complexity growth of operators in
  the {SYK} model and in {JT} gravity},
  \href{https://doi.org/10.1007/jhep03(2021)014}{\emph{Journal of High Energy
  Physics} {\bfseries 2021} (2021) 14}
  [\href{https://arxiv.org/abs/2008.12274}{{\ttfamily 2008.12274}}].

\bibitem{lensky2020}
Y.~D.~Lensky, X.-L.~Qi and P.~Zhang, \emph{Size of bulk fermions in the {SYK}
  model}, \href{https://doi.org/10.1007/jhep10(2020)053}{\emph{Journal of High
  Energy Physics} {\bfseries 2020} (2020) 53}
  [\href{https://arxiv.org/abs/2002.01961}{{\ttfamily 2002.01961}}].

\bibitem{qi2019}
X.-L.~Qi and A.~Streicher, \emph{Quantum epidemiology: Operator growth, thermal
  effects, and {SYK}},
  \href{https://doi.org/10.1007/jhep08(2019)012}{\emph{Journal of High Energy
  Physics} {\bfseries 2019} (2019) 12}
  [\href{https://arxiv.org/abs/1810.11958}{{\ttfamily 1810.11958}}].

\bibitem{lucas2020}
A.~Lucas, \emph{Non-perturbative dynamics of the operator size distribution in
  the {S}achdev–{Y}e–{K}itaev model},
  \href{https://doi.org/10.1063/1.5133964}{\emph{Journal of Mathematical
  Physics} {\bfseries 61} (2020) 081901}
  [\href{https://arxiv.org/abs/1910.09539}{{\ttfamily 1910.09539}}].

\bibitem{lin2019}
H.~W.~Lin, J.~Maldacena and Y.~Zhao, \emph{Symmetries near the horizon},
  \href{https://doi.org/10.1007/jhep08(2019)049}{\emph{Journal of High Energy
  Physics} {\bfseries 2019} (2019) 49}
  [\href{https://arxiv.org/abs/1904.12820}{{\ttfamily 1904.12820}}].

\bibitem{groszkowski2022}
P.~Groszkowski, A.~Seif, J.~Koch and A.~A.~Clerk, \emph{Simple master equations
  for describing driven systems subject to classical non-{M}arkovian noise},
  {\emph{ArXiv} (2022) } [\href{https://arxiv.org/abs/2207.03980}{{\ttfamily
  2207.03980}}].

\bibitem{torlai2020}
G.~Torlai, G.~Mazzola, G.~Carleo and A.~Mezzacapo, \emph{Precise measurement of
  quantum observables with neural-network estimators},
  \href{https://doi.org/10.1103/PhysRevResearch.2.022060}{\emph{Phys. Rev.
  Research} {\bfseries 2} (2020) 022060}
  [\href{https://arxiv.org/abs/1910.07596}{{\ttfamily 1910.07596}}].

\end{thebibliography}\endgroup

\end{document}